\hfuzz 2pt
\font\titlefont=cmbx10 scaled\magstep1
\magnification=\magstep1

\null
\vskip 2cm
\centerline{\titlefont DISSIPATIVE EFFECTS}
\smallskip
\centerline{\titlefont IN SEMILEPTONIC B-$\overline{\hbox{B}}$ DECAYS}
\vskip 2.5cm
\centerline{\bf F. Benatti}
\smallskip
\centerline{Dipartimento di Fisica Teorica, Universit\`a di Trieste}
\centerline{Strada Costiera 11, 34014 Trieste, Italy}
\centerline{and}
\centerline{Istituto Nazionale di Fisica Nucleare, Sezione di 
Trieste}
\vskip 1cm
\centerline{\bf R. Floreanini}
\smallskip
\centerline{Istituto Nazionale di Fisica Nucleare, Sezione di 
Trieste}
\centerline{Dipartimento di Fisica Teorica, Universit\`a di Trieste}
\centerline{Strada Costiera 11, 34014 Trieste, Italy}
\vskip 2cm
\centerline{\bf Abstract}
\smallskip
\midinsert
\narrower\narrower\noindent
We study the time evolution and decay of the neutral $B$-meson system using
quantum dynamical semigroups. New, nonstandard terms appear in
the expression of relevant observables; they can be parametrized
in terms of six phenomenological constants. Although very small, these
parameters may be detected in the new generation of dedicated $B$-meson
experiments.
\endinsert
\bigskip
\vfil\eject

Dynamics that extend the time evolution predicted
by ordinary quantum mechanics have been studied in various physical
phenomena. Typical examples are the open quantum systems,
represented as small subsystems in interaction with large environments.
Although the total system is closed and its time evolution is standard
and unitary, the dynamics of the subsystem, obtained by eliminating the
environment degrees of freedom, develops dissipation and irreversibility.

Assuming the interaction between subsystem and environment to be ``weak'',
a condition met in most physical situations, the sub-dynamics
can be given a precise mathematical description in terms of the so-called
quantum dynamical semigroups. These are entropy-increasing, linear maps that
transform density matrices into density matrices ({\it i.e.} state into
states), with forward in time composition law (semigroup property), that
satisfy the condition of complete positivity.[1-3]

This description is very general; although developed in the framework of
quantum optics,[4-6] it has been successfully applied to model a large variety
of different physical situations. In particular, quantum dynamical semigroups
have been used in the study of unstable systems.[7-9] It is therefore natural
to look for signals of induced dissipative effects in the dynamics and 
decay of elementary particles. Indeed, general considerations based on
quantum gravity effects [10-15] and the dynamics of extended objects [16, 17]
suggests that completely positive time evolutions can be expected in the
description of many particle physics phenomena, as the result of the 
interaction with a ``stringy'' environment.

In this respect, particularly suitable for the analysis of dissipative,
non-standard phenomena are neutron interferometers[11, 18]
and systems of neutral mesons.[19-24] 
Detailed study of the $K^0$-$\overline{K^0}$
system using quantum dynamical semigroups has been performed, in the case
of both single and correlated kaons;[21-24] the outcome of 
these investigations is that the
dissipative contribution in the extended dynamics could be in the
reach of the next generation of kaon experiments.

In the following, we shall devote our attention to the 
$B^0$-$\overline{B^0}$ system and analyze its time-evolution using
a similar approach. Dedicated $B$-meson experiments
are now collecting data, so that it appears timely to discuss possible
signals of dissipative effects in the observables these experiments
will measure.

We shall study semileptonic decays of both single and correlated
neutral $B$-mesons and show that certain combinations of observables
directly accessible to the experiment are particularly sensitive to
the dissipative effects, quite independently from the values of the
other standard $CP$ and $CPT$ violating phenomenological parameters.%
\footnote{$^\dagger$}{For recent studies on possible violations of the
$CPT$ symmetry in $B$-decays, see [25-28] and references therein.}
Although more complete studies, that include precise analysis of acceptance
and efficiency of the various detectors, are certainly necessary,
our investigation suggests that
future $B$-meson experiments should be able to put stringent
limits on the parameters that describe non-standard, dissipative phenomena.

\bigskip

As in the case of the $K^0$-$\overline{K^0}$ system, the evolution and decay
of the neutral $B$-meson system can be effectively modelled by means
of a two-dimensional Hilbert space. (We shall limit our discussion to the
$B_d$-mesons, although most discussions below apply to the $B_s$-mesons
as well.) The states of this system will be
described by a density matrix $\rho$. 
In the basis $|B^0\rangle$, $|\overline{B^0}\rangle$,
it can be written as:
$$
\rho=\left(\matrix{
\rho_1&\rho_3\cr
\rho_4&\rho_2}\right)\ , \eqno(1)
$$
where $\rho_4\equiv\rho_3^*$, and $*$ signifies complex conjugation.

The evolution in time of this matrix can be described quite in general
by an equation of the following form:
$$
{\partial\rho(t)\over\partial t}=-iH\, \rho(t)+i\rho(t)\, H^\dagger 
+L[\rho] .\eqno(2)
$$
The first two pieces in the r.h.s. give the standard hamiltonian
contribution, while $L$ is a linear map that encodes possible dissipative,
non-standard effects.

The effective hamiltonian $H$ includes a non-hermitian part,
$H=M-{i\over 2}{\mit\Gamma}$, that characterizes the natural width of the states.
The entries of $H$ can be expressed in terms of its eigenvalues:
$\lambda_S=m_S-{i\over 2}\gamma_S$, $\lambda_L=m_L-{i\over 2}\gamma_L$,
and the two complex parameters $\epsilon_S$, $\epsilon_L$, appearing in the
corresponding eigenstates, 
$$
|B_S\rangle={1\over\sqrt{2\,(1+|\epsilon_S|^2)}}
\left(\matrix{1+\epsilon_S\cr1-\epsilon_S}\right)\ ,\quad
|B_L\rangle={1\over\sqrt{2\,(1+|\epsilon_L|^2)}}
\left(\matrix{\phantom{-}1+\epsilon_L\cr -1+\epsilon_L}\right)\ .\eqno(3)
$$
We use here a notation that follows the conventions adopted for the
neutral kaon system.[25] The two states in (3) are expected to have a 
negligible width difference,
$$
\Delta\Gamma\ll \Gamma\ ,\qquad \Delta\Gamma=\gamma_S-\gamma_L\ ,
\qquad \Gamma={\gamma_S+\gamma_L\over2}\ ,\eqno(4)
$$
so that they are more conveniently distinguished via their mass difference,
$\Delta m=m_L - m_S$,
rather than their different lifetimes. The notation $|B_L\rangle$, for the
light meson, and $|B_H\rangle$, for the heavy partner are therefore 
sometimes used instead of those in (3).

The form of the additional piece $L[\rho]$ in (2)
is fixed by the physical requirements that the complete
time-evolution, $\gamma_t: \rho(0)\mapsto\rho(t)$, 
needs to satisfy. Generally, the one parameter (=time) family of linear maps
$\gamma_t$ should transform density matrices into density matrices and
have the properties of increasing the entropy, of obeying the semigroup
composition law, $\gamma_t[\rho(t')]=\rho(t+t')$, for $t,\ t'\geq0$,
of preserving the positivity of $\rho(t)$ for all times.
Actually, for the physical consistency of the formalism,
in the case of correlated systems, one has to demand that
the time evolution $\gamma_t$ be completely positive.[24]

With these requirements, the linear map $L[\rho]$ can be parametrized
in terms of six real constants. To write it down explicitly, one expands
the density matrix $\rho$ in terms of Pauli matrices $\sigma_i$, $i=1,2,3$,
and the unit matrix $\sigma_0$: 
$\rho=\rho_\mu\, \sigma_\mu$, $\mu=\,0$, 1, 2, 3; then, the map $L$ can
be represented by a symmetric $4\times 4$ matrix $\big[L_{\mu\nu}\big]$, 
acting on the vector with components $(\rho_0,\rho_1,\rho_2,\rho_3)$.
Explicitly one finds:[21]
$$
\big[L_{\mu\nu}\big]=-2\left(\matrix{0&0&0&0\cr
                                     0&a&b&c\cr
                                     0&b&\alpha&\beta\cr
                                     0&c&\beta&\gamma\cr}\right)
\ ,\eqno(5)
$$
with $a$, $\alpha$ and $\gamma$ non-negative. The parameters
$a$, $b$, $c$, $\alpha$, $\beta$, and $\gamma$ are not all independent; 
to assure the complete positivity
of the time-evolution generated by (2), they have to satisfy
certain inequalities (for details, see [21, 23]).

Physical observables of the neutral $B$-meson system 
can be obtained from the density
matrix $\rho(t)$ obeying (2) by taking its trace with suitable hermitian
operators. In order to study the time evolution of these observables, one has to
solve the equation (2) for an arbitrary initial state $\rho(0)$.

Rough dimensional estimates based on phenomenological
considerations predict the non-standard parameters to be very small,
at most of order $m_B^2/M_P\approx 10^{-18}\ {\rm GeV}$, where $m_B$ is the
$B$-meson mass and $M_P$ is Planck's mass.[19-21] The piece $L[\rho]$ in (2)
can then be treated as a perturbation. Furthermore, to simplify the
computations, we shall work in a phase convention for the states
$|B^0\rangle$, $|\overline{B^0}\rangle$ such that the $CP$ violation
induced by the effective hamiltonian $H$ is small, of order $10^{-3}$
(in practice, this amounts assuming ${\cal I}m(\epsilon_S +\epsilon_L)$
to be of the same order of magnitude). As a consequence, possible $T$
and $CPT$ violating effects in $H$ are also small.[25, 27] 
With this phase choice,
the quantities $\epsilon_S\Delta\Gamma$ and 
$\epsilon_L\Delta\Gamma$ turn out to be roughly of the same order of
magnitude of the non-standard parameters $a$, $b$, $c$, $\alpha$, 
$\beta$, and $\gamma$. The use of perturbation theory in all these
quantities is therefore justified.

It should be noticed that the above phase choice differs from the phase
convention usually adopted in the Standard Model for the study of the
$B^0$-$\overline{B^0}$ system (which implies a sizable imaginary part
for the combination $\epsilon_S + \epsilon_L$). 
In the present work, we focus our attention on the dissipative part
$L[\rho]$ of the evolution equation (2), parametrized as in (5) in terms
of $a$, $b$, $c$, $\alpha$, $\beta$, and $\gamma$.
Since the map $L$ is linear, these parameters, being small in one phase
convention, remain small in any other phase choice.
Therefore, for our purposes, the rather unusual phase convention adopted
in the following%
\footnote{$^\dagger$}{Note that precisely this
phase convention has been used to present some measured data by 
the OPAL Collaboration [29, 30].}
does not affect in any way the generality of our discussion.

As mentioned in the introductory remarks, we shall limit our analysis to the
study of semileptonic decays $h\ell\nu$ of the neutral $B$-meson system,
where $h$ stands for any allowed charged hadronic state.
We shall be as general as possible, and include in our discussion
possible violations of the $\Delta B=\Delta Q$ rule.
The amplitudes for the decay of a $B^0$ or $\overline{B^0}$ state
into $h^-\ell^+\nu$ and $h^+\ell^-\bar\nu$ can then be parametrized
in terms of three complex constants, $x_h$, 
$y_h$ and $z_h$, as follows:
$$
\eqalignno{
&{\cal A}(B^0\rightarrow h^-\ell^+\nu)={\cal M}_h (1-y_h)\ , &(6a)\cr
&{\cal A}(\overline{B^0}\rightarrow h^+\ell^-\bar\nu)=
{\cal M}_h^* (1+y^*_h)\ , &(6b)\cr
&{\cal A}(B^0\rightarrow h^+\ell^-\bar\nu)= z_h\, 
{\cal A}(\overline{B^0}\rightarrow h^+\ell^-\bar\nu)\ , &(6c)\cr
&{\cal A}(\overline{B^0}\rightarrow h^-\ell^+\nu)=
x_h\, {\cal A}(B^0\rightarrow h^-\ell^+\nu)\ , &(6d) }
$$
where ${\cal M}_h$ is a common factor.
(Sometimes,[31] $\bar x_h\equiv z_h^*$ is used instead of $z_h$.)
The $\Delta B=\Delta Q$ rule requires $x_h=z_h=\,0$, while
$CPT$-invariance, $y_h=\,0$.

The hermitian operators that describe these semileptonic decays can be easily
obtained from (6); explicitly, in the $|B^0\rangle$, $|\overline{B^0}\rangle$
basis, one gets:
$$
{\cal O}_{h^-}={|{\cal M}_h|^2\over2}\,
|1-y_h|^2\ \left[\matrix{1&x_h\cr
                         x_h^*&|x_h|^2\cr}\right]\ ,\qquad
{\cal O}_{h^+}={|{\cal M}_h|^2\over2}\,
|1+y_h|^2\ \left[\matrix{|z_h|^2&z_h^*\cr
                          z_h&1\cr}\right]\ .\eqno(7)
$$
The quantities $x_h$, $y_h$ and $z_h$ are expected to be very small;
therefore, in the following we shall treat them as the other small
parameters in the theory.
In the computation of semileptonic observables below, we shall
keep only first order terms in all these parameters; this approximation
turns out to be more than adequate for our considerations.

We shall first study observables connected to the time-evolution and decay
of a single, uncorrelated $B^0$-$\overline{B^0}$ system, that can be typically
measured at colliders. Let us indicate with $\rho_{B^0}(t)$,
$\rho_{\bar B^0}(t)$ the time evolution according to (2) of a state
that is initially a pure $B^0$, $\overline{B^0}$ meson, described
by the density matrices:
$$\rho_{B^0}=\left[\matrix{1&0\cr 
                           0&0\cr}\right]\ ,\qquad
  \rho_{\bar B^0}=\left[\matrix{0&0\cr 
                           0&1\cr}\right]\ .\eqno(8)
$$
The probability rate that an initial $\rho_{B^0}$, $\rho_{\bar B^0}$
state decays at time $t$ into a given semileptonic state $h\ell\nu$ described
by one of the operators in (7) is then given by:
$$
{\cal P}_h(B^0;t)={\rm Tr}\big[\rho_{B^0}(t)\, {\cal O}_h\big]\ ,\qquad
{\cal P}_h(\overline{B^0};t)={\rm Tr}\big[\rho_{\bar B^0}(t)\, {\cal O}_h\big]\ .
\eqno(9)
$$
In writing down the explicit form of the probabilities $\cal P$, it is convenient
to introduce the new variable $\tau=t\,\Gamma$; in practice, $\tau$ expresses the
time variable in units of the $B$ lifetime. It is also customary to define
the two combinations:%
\footnote{$^\dagger$}{They are usually called $x_B$ and $y_B$, respectively; we
do not use these labels to avoid confusion with the parameters introduced
in (6).}
$$
\omega={\Delta m\over\Gamma}\ ,\qquad\qquad \delta={\Delta\Gamma\over 2\Gamma}\ .
\eqno(10)
$$
Although not directly measured, $\delta$ is expected to be very small,
$\delta\leq 10^{-2}$, while the most recent data give:
$\omega=0.734\pm0.035$.[32] In discussing $B^0$-$\overline{B^0}$ observables, 
one often takes the simplified assumption $\delta=0$; in the following, we shall
keep $\delta$ nonvanishing, unless explicitly stated.

Up to first order in all small parameters, the probabilities (9), apart from 
an overall exponential decay, have oscillatory behaviour modulated
by $\omega$ and exponential behaviour dictated by $\delta$:
$$
\eqalignno{
{\cal P}_{h^+}(B^0;\tau)={|{\cal M}_h|^2\over2}e^{-\tau}\Big\{
&\cos\omega\tau\Big[2\,{\cal R}e(\epsilon_S+\epsilon_L)-2\,{\cal R}e(y_h)
-e^{-(A-D)\tau}\Big]\cr
+&\sin\omega\tau\Big[2\,{\cal I}m(z_h)+{4\,\delta\over\delta^2+\omega^2}
{\cal I}m(C)-{\cal R}e(B)\Big]\cr
+&\cosh\delta\tau\Big[1-2\,{\cal R}e(\epsilon_S+\epsilon_L)
+2\,{\cal R}e(y_h)\Big]\cr
-&\sinh\delta\tau\Big[2\,{\cal R}e(z_h)-{D\over\delta}
+{4\,\omega\over\delta^2+\omega^2}{\cal I}m(C)\Big]\Big\}\ ,&(11a)\cr
&&\cr
{\cal P}_{h^-}(B^0;\tau)={|{\cal M}_h|^2\over2}e^{-\tau}\Big\{
-&\cos\omega\tau\Big[2\,{\cal R}e(y_h)-e^{-(A-D)\tau}
+{4\,\delta\over\delta^2+\omega^2}{\cal R}e(C)\Big]\cr
-&\sin\omega\tau\Big[2\,{\cal I}m(\epsilon_S-\epsilon_L)
+2\,{\cal I}m(x_h)-{\cal R}e(B)\Big]\cr
+&\cosh\delta\tau\Big[1-2\,{\cal R}e(y_h)
+{4\,\delta\over\delta^2+\omega^2}{\cal R}e(C)\Big]\cr
-&\sinh\delta\tau\Big[2\,{\cal R}e(\epsilon_S-\epsilon_L)
+2\,{\cal R}e(x_h)-{D\over\delta}\Big]\Big\}\ ,&(11b)\cr
}
$$
where the convenient shorthand notations have been used:
$$
A={a+\alpha\over\Gamma}\ ,\quad
B={\alpha-a+2ib\over\Delta m}\ ,\quad
C={c+i\beta\over\Gamma}\ ,\quad
D={\gamma\over\Gamma}\ ;\eqno(12)
$$
the expressions for ${\cal P}_{h^+}(\overline{B^0};\tau)$ and 
${\cal P}_{h^-}(\overline{B^0};\tau)$ are obtained from $(11b)$ and $(11a)$
by changing the sign of $\epsilon_S$, $\epsilon_L$, $y_h$, $C$ and 
exchanging $x_h$ and $z_h$.

The decay rates $\cal P$ can be observed at colliders;
in particular, the analysis reported in Ref.[29] for the standard case
can surely be repeated with the extended expressions above.
Nevertheless, it is more convenient to construct specific asymmetries
by taking suitable combinations of the $\cal P$'s. Two independent asymmetries
that can then be constructed are:
$$
\eqalignno{
&A_T(\tau)={ {\cal P}_{h^-}(\overline{B^0};\tau) -
             {\cal P}_{h^+}(B^0;\tau)\over
             {\cal P}_{h^-}(\overline{B^0};\tau)+
			 {\cal P}_{h^+}(B^0;\tau)}\ , &(13a)\cr
&A_{CPT}(\tau)={   {\cal P}_{h^+}(\overline{B^0};\tau) -
                   {\cal P}_{h^-}(B^0;\tau)\over
                   {\cal P}_{h^+}(\overline{B^0};\tau) +
                   {\cal P}_{h^-}(B^0;\tau)}\ . &(13b)}
$$
The first one signals the violation of time reversal by measuring
the rate difference between the process $B^0\rightarrow\overline{B^0}$
and its time-conjugate $\overline{B^0}\rightarrow B^0$,
while the second asymmetry is sensitive to $CPT$ violating effects,
by comparing the rate difference between the process
$B^0\rightarrow B^0$ and its $CPT$-conjugate
$\overline{B^0}\rightarrow\overline{B^0}$.
Using the expressions for the probabilities $\cal P$ given before,
one explicitly finds (for $\tau$ not too small, 
in the case of the first asymmetry):
$$
\eqalignno{
&A_T(\tau)=2\,{\cal R}e(\epsilon_S+\epsilon_L)+2\,{\cal R}e(y_h)\cr
&\hskip 2.5cm+{1\over\cosh\delta\tau-\cos\omega\tau}\bigg[
\sin\omega\tau\bigg({\cal I}m(x_h-z_h)
-{4\,\delta\over\delta^2+\omega^2}{\cal I}m(C)\bigg)\cr
&\hskip 2.5cm+\sinh\delta\tau\bigg({\cal R}e(z_h-x_h)+
{4\,\omega\over\delta^2+\omega^2}{\cal I}m(C)\bigg)\bigg]\ , &(14a)\cr
&&\cr
&A_{CPT}(\tau)=2\,{\cal R}e(y_h)
-{4\,\delta\over\delta^2+\omega^2}
\bigg({\cosh\delta\tau-\cos\omega\tau\over
\cosh\delta\tau+\cos\omega\tau}\bigg)\,{\cal R}e(C)\cr
&\hskip 2.5cm+{1\over\cosh\delta\tau+\cos\omega\tau}\bigg[
\sin\omega\tau\bigg(2\,{\cal I}m(\epsilon_S-\epsilon_L)+{\cal I}m(x_h-z_h) 
\bigg)\cr
&\hskip 2.5cm+\sinh\delta\tau\bigg(2\,{\cal R}e(\epsilon_S-\epsilon_L)
+{\cal R}e(x_h-z_h)
\bigg)\bigg]\ . &(14b)\cr
}
$$

Other independent asymmetries can be constructed with the semileptonic
decay rates $\cal P$; they all involve the non-standard parameter $C$
besides $\epsilon_S$, $\epsilon_L$, $x_h$, $y_h$, $z_h$. Therefore,
by fitting the experimental data with the explicit expression of all
these asymmetries one should be able to extract estimates for $C$.
Unfortunately, as apparent from (14), $C$ is always accompanied by a factor
$\delta$, so that its dependence is further suppressed with respect to the
other small parameters. In particular,
the approximation $\delta\approx 0$ would completely eliminate the
presence of this parameter from (14).
The asymmetries $A_T$ and $A_{CPT}$ (and analogue ones) are therefore
not suitable for identifying dissipative effects; instead, they can be used
to obtain estimates on some combinations of the conventional $CP$
and $CPT$ violating parameters also in presence of dissipative phenomena.

Nevertheless, one can construct more complicated combinations of
the semileptonic probability rates $\cal P$, which are particularly
sensitive to the non-standard parameters. One illuminating example is
given by:
$$
A_{\Delta m}(\tau)={ \big[{\cal P}_{h^-}(B^0;\tau)-
{\cal P}_{h^-}(\overline{B^0};\tau)\big]
- \big[{\cal P}_{h^+}(B^0;\tau)- {\cal P}_{h^+}(\overline{B^0};\tau)\big]
\over
{\cal P}_{h^-}(B^0;\tau)+{\cal P}_{h^-}(\overline{B^0};\tau)
+{\cal P}_{h^+}(B^0;\tau)+ {\cal P}_{h^+}(\overline{B^0};\tau)}\ .
\eqno(15)
$$
In the approximation $\delta\approx0$, it assumes a very
simple form:
$$
A_{\Delta m}(\tau)=e^{-A\tau}\, \cos\omega\tau+\sin\omega\tau\big[
{\cal R}e(B) - {\cal I}m(x_h+z_h)\big]\ .\eqno(16)
$$
Thanks to the different time-behaviour, a measure of $A_{\Delta m}$
allows, at least in principle, a determination of the parameter $A$
and of the combination ${\cal R}e(B) - {\cal I}m(x_h+z_h)$;
assuming the validity of the $\Delta B=\Delta Q$ rule, via the definitions
in (12), one can then obtain an estimate of the two dissipative
parameters $a$ and $\alpha$. The actual accuracy of such a determination 
highly depends on the sensitivity of the measure of $A_{\Delta m}$.
Although a specific discussion on this point is surely beyond the scope
of the present work, simulations presented in [25] in the case of similar
observables suggest that the accuracy on the measure of (15) at present
colliders can be estimated in about ten percent; this sensitivity is already
enough to give interesting bounds on $a$ and $\alpha$. However, future 
dedicated $B$-experiments can greatly improve these estimates.

Other useful asymmetries can be obtained by considering 
time-integrated rates:
$$
{\cal P}_h(B)={1\over\Gamma}\int_0^\infty d\tau\ {\cal P}_h(B;\tau)\ .
\eqno(17)
$$
Using again a combination analogous to the one in (15), one gets,
in the same approximation:
$$
A_{\Delta m}'={1\over 1+\omega^2}\bigg\{1-\omega\, {\cal I}m(x_h+z_h) 
+ \omega\, {\cal R}e(B) + {1\over 1+\omega^2}\Big[(\omega^2-1)A
-2\,\omega^2D\Big]\bigg\}\ .\eqno(18)
$$
Assuming again the $\Delta B=\Delta Q$ rule, 
a measure of $A'_{\Delta m}\, (1+\omega^2)$ not compatible with 1
would clearly signal the presence
of non-standard effects; this is surely
the most simple check on the extended dynamics given by (2)
that can be performed at colliders, using semileptonic decays.

Further insights on the dissipative phenomena described
by the non-standard time evolution generated by (2) can be
obtained from experiments at the so-called $B$-factories;
indeed, correlated $B^0$-$\overline{B^0}$ systems look particularly suitable
for studying phenomena involving loss of quantum coherence.
In those set-ups, correlated $B^0$-$\overline{B^0}$ mesons are produced from
the decay of the $\Upsilon(4S)$ resonance.
Since the $\Upsilon$-meson has spin 1, its decay into two spinless bosons produces
an antisymmetric spatial state. In the $\Upsilon$ rest frame, the two neutral 
$B$-mesons are produced flying apart with opposite momenta; 
in the basis $|B^0\rangle$,
$|\overline{B^0}\rangle$, the resulting state can be described by:
$$
|\psi_A\rangle= {1\over\sqrt2}\Big(|B^0,-p\rangle \otimes  
|\overline{B^0},p\rangle -
|\overline{B^0},-p\rangle \otimes  |B^0,p\rangle\Big)\ .\eqno(19)
$$
The corresponding density operator $\rho_A$ is a $4\times 4$ matrix
that can be conveniently written in terms of single kaon projectors:
$$
P_1\equiv |B^0\rangle\langle B^0|=
    \left(\matrix{1 & 0\cr 0 & 0\cr}\right)\ ,\qquad
P_2\equiv |\overline{B^0}\rangle\langle \overline{B^0}|=
    \left(\matrix{0 & 0\cr 0 & 1\cr}\right)\ ,\eqno(20a)
$$
and the off-diagonal operators
$$
P_3\equiv |B^0\rangle\langle \overline{B^0}|=
    \left(\matrix{0 & 1\cr 0 & 0\cr}\right)\ ,\qquad
P_4\equiv |\overline{B^0}\rangle\langle B^0|=
    \left(\matrix{0 & 0\cr 1 & 0\cr}\right)\ .\eqno(20b)
$$
Explicitly, one finds:
$$
\rho_A={1\over 2}\Big[P_1\otimes P_2\ +\ P_2\otimes P_1\ -\ 
P_3\otimes P_4\ -\ P_4\otimes P_3\Big]\ .\eqno(21)
$$

The time evolution of a system of two correlated neutral $B$-mesons,
initially described by $\rho_A$, can be analyzed using the single
$B$-meson dynamics previously discussed.
In studying the time evolution of $\rho_A$, we shall assume that, once
produced in a $\Upsilon$ decay, the kaons evolve in time each according to the
completely positive map $\gamma_t$ generated by (2).
This assures the positivity of the eigenvalues of any physical
states at all times.[24]

Within this framework, the density matrix that describes a situation in which
the first $B$-meson has evolved up to proper time $t_1$ 
and the second up to proper time $t_2$ is given by:
$$
\eqalign{
\rho_A(t_1&,t_2)\equiv
\big(\gamma_{t_1}\otimes\gamma_{t_2}\big)\big[\rho_A\big]\cr
=&{1\over 2}\Big[P_1(t_1)\otimes P_2(t_2)\ 
+\ P_2(t_1)\otimes P_1(t_2)\ 
- P_3(t_1)\otimes P_4(t_2)-P_4(t_1)\otimes P_3(t_2)\Big]\ ,}
\eqno(22)
$$
where $P_i(t_1)$ and $P_i(t_2)$, $i=1,2,3,4$, represent the evolution
according to (2) of the initial operators (20), up to the time $t_1$
and $t_2$, respectively.

The typical observables that can be studied with correlated mesons systems 
are double decay rates, {\it i.e.} the probabilities 
${\cal G}(f_1,t_1; f_2,t_2)$ that a meson decays
into a final state $f_1$ at proper time $t_1$, while the other meson
decays into the final state $f_2$ at proper time $t_2$.
They can be computed using:
$$
{\cal G}(f_1,t_1; f_2,t_2)=
\hbox{Tr}\Big[\Big({\cal O}_{f_1}\otimes{\cal O}_{f_2}\Big) 
\rho_A(t_1,t_2)\Big]\ ,\eqno(23)
$$
where ${\cal O}_{f_1}$, ${\cal O}_{f_2}$ represent $2\times2$ hermitian
matrices describing the decay of a single meson into the final
states $f_1$, $f_2$, respectively. For the purpose of the present note,
that deals with semileptonic decays, they can be identified with one
of the matrices in (7).

Unfortunately, the $B$-mesons decay too quickly to allow in general
a precise enough measure of the double decay rates in (23).
Therefore, much of the analysis at $B$-factories is carried out 
using integrated distributions at fixed time interval $t=t_1-t_2$.
One then deals with single-time distributions, defined by
$$
{\mit\Gamma}(f_1,f_2;t)=\int_0^\infty dt'\, {\cal G}(f_1,t'+t;f_2,t')\ ,
\eqno(24)
$$
where $t$ is taken to be positive. For negative $t$, one defines:
$$
{\mit\Gamma}(f_1,f_2;-|t|)=\int_0^\infty dt'\, {\cal G}(f_1,t'-|t|;f_2,t')\ 
\theta(t'-|t|)\ ;
\eqno(25)
$$
the presence of the step-function is necessary since the evolution is
of semigroup type, with forward in time propagation, starting from zero
(we can not propagate a kaon before it is created in a $\Upsilon$-decay).
In this case, one easily finds: ${\mit\Gamma}(f_1,f_2;-|t|)=
{\mit\Gamma}(f_2,f_1;|t|)$. In the following, we shall always assume:
$t\geq0$.

The dynamics generated by (2) gives results for the double probabilities
${\mit\Gamma}(f_2,f_1;t)$ that substantially differ from the ones obtained
using ordinary quantum mechanics. The most striking difference arises when
the final states coincide $f_1=f_2=f$ and $t$ becomes small. Due to the
antisymmetry of the initial state $\rho_A$ in (21), quantum mechanics
predicts a vanishing value for ${\mit\Gamma}(f,f;0)$, while in general
this is not the case for the completely positive dynamics generated
by (2). The quantities ${\mit\Gamma}(f,f;0)$ are therefore very sensitive
to the dissipative parameters $a$, $b$, $c$, $\alpha$, $\beta$ and $\gamma$.

Indeed, in the case of semileptonic final states, 
neglecting higher order terms in $\delta$ multiplying small parameters,
one finds:
$$
\eqalign{
{\mit\Gamma}(h^\pm,h^\pm;0)=&{|{\cal M}_h|^4\over 8\Gamma}\bigg\{
A-{\omega\over 1+\omega^2}\Big[{\cal R}e(B)-\omega\, {\cal I}m(B)\Big]\cr
&\hskip 2cm\pm {8\delta\over(4+\omega^2)^2}
\Big[ (4-\omega^2)\, {\cal R}e(C)
+4\,\omega\, {\cal I}m(C)\Big]\bigg\}\ .}
\eqno(26)
$$
It turns out that also the remaining two semileptonic probabilities at $t=\,0$
depend only on the non-standard parameters:
$$
{\mit\Gamma}(h^\pm,h^\mp;0)={|{\cal M}_h|^4\over 8\Gamma}\bigg\{
2(1+D)-A+{\omega\over 1+\omega^2}\Big[{\cal R}e(B) - \omega\, {\cal I}m(B)
\Big]\bigg\}\ .\eqno(27)
$$
More in general, one can study the explicit time-dependence of various
semileptonic asymmetries that can be constructed out of the integrated
probabilities (24). Using again the variable $\tau=t\,\Gamma$, the analogous
of the asymmetry $(13a)$,
$$
{\cal A}_T(\tau)={ {\mit\Gamma}(h^-,h^-;\tau)-{\mit\Gamma}(h^+,h^+;\tau)
\over
{\mit\Gamma}(h^+,h^-;\tau)+{\mit\Gamma}(h^-,h^+;\tau)}\ ,\eqno(28)
$$
takes the following explicit expression (up to first order in $\delta$ 
and for $\tau$ not too large):
$$
\eqalign{
{\cal A}_T(\tau)=&{2\over1+\cos\omega\tau}\bigg\{
\big[{\cal R}e(\epsilon_S+\epsilon_L) -2\,{\cal R}e(y_h)\big]
(1-\cos\omega\tau)\cr
+&{2\delta\over(4+\omega^2)^2}\Big[\big[(\omega^2-4)\, {\cal R}e(C)
-4\,\omega\, {\cal I}m(C)\big](1+\cos\omega\tau)\cr
-&{4\over\omega^2}\big[(4+3\omega^2)\, {\cal I}m(C)
-\omega^3\, {\cal R}e(C)\big]\sin\omega\tau
-{2(4+\omega^2)\over\omega} \big[\omega\, {\cal R}e(C)-2\, {\cal I}m(C)\big]
\tau\Big]\bigg\}\ .}
\eqno(29)
$$
Here again one notices the suppression of the $C$-dependence 
by a factor $\delta$ with respect to the other small parameters:
it is therefore difficult to extract informations
about $C$ by fitting (29) with the experimental data.
However, as $\tau$ becomes small, the expression for ${\cal A}_T$ simplifies;
in particular, as already noted, the dependence on $\epsilon_S$, $\epsilon_L$
and $y_h$ disappear:
$$
{\cal A}_T(0)={4\, \delta\over(4+\omega^2)^2}\Big[
(\omega^2-4)\, {\cal R}e(C)-4\,\omega\, {\cal I}m(C)\Big]\ .\eqno(30)
$$
A non-vanishing measure of ${\cal A}_T(0)$ would therefore provide
evidence for $C$. 

Similarly, in the case of the combination
$$
{\cal B}(\tau)={ {\mit\Gamma}(h^+,h^+;\tau)+{\mit\Gamma}(h^-,h^-;\tau)
\over
{\mit\Gamma}(h^+,h^+;\tau)+{\mit\Gamma}(h^+,h^-;\tau)}\ ,\eqno(31)
$$
one finds, neglecting terms proportional to $\delta$ times $C$ in the
denominator,
$$
{\cal B}(0)={\Lambda\over 1+D}\ ,\eqno(32)
$$
where
$$
\Lambda=A+{\omega\over1+\omega^2}\big[\omega\, {\cal I}m(B)-{\cal R}e(B)\big]\ .
\eqno(33)
$$
This combination of non-standard parameters appears also in the study
of the asymmetry ${\cal A}_{\Delta m}(\tau)$, analogous to the one in (15)
for single-meson systems:
$$
{\cal A}_{\Delta m}(\tau)={ \big[{\mit\Gamma}(h^+,h^+;\tau)
+{\mit\Gamma}(h^-,h^-;\tau)\big] - 
\big[{\mit\Gamma}(h^+,h^-;\tau)
+{\mit\Gamma}(h^-,h^+;\tau)\big]
\over
{\mit\Gamma}(h^+,h^+;\tau) +{\mit\Gamma}(h^-,h^-;\tau)
+{\mit\Gamma}(h^+,h^-;\tau) + {\mit\Gamma}(h^-,h^+;\tau)}\ .
\eqno(34)
$$
It takes a very simple form (we neglect terms containing
$\delta$ times small parameters):
$$
{\cal A}_{\Delta m}(\tau)=\Big[A(1+\tau)+{\cal I}m(B)-1\Big]\cos\omega\tau
+{ {\cal R}e\big[B(i-\omega)e^{i\omega\tau}\big] \over 1+\omega^2}\ ,
\eqno(35)
$$
so that,
$$
{\cal A}_{\Delta m}(0)= \Lambda-1\ .\eqno(36)
$$
Once the combination $\Lambda$ is measured via (36), one can deduce the
value of $D$ from (32) and a measure of ${\cal B}(0)$.
Further, by fitting experimental data with (35), from the different
time behaviours one can extract $A$ and the combination
$\omega\, {\cal I}m(B)-{\cal R}e(B)$.

The asymmetry in (34) will be accurately measured at the $B$-factories,
since it is used to obtain a precise determination of the mass
difference $\Delta m$.[33, 34] An independent analysis of the form (35)
of ${\cal A}_{\Delta m}$ is certainly needed in order to estimate precisely
the sensitivity of those experiments to the dissipative parameters $A$
and the combination $\omega\, {\cal I}m(B)-{\cal R}e(B)$. However,
from the discussion in [33] and the simulations presented in [27],
in which a simplified version of the extended dynamics (2), (5) is used,
one can reasonably expect that the actual measured data will at the end
constrain the dissipative terms, as well as the standard $CP$ and $CPT$
violating parameters, to a few percent level. This would allow to establish
significant limits on the presence of dissipative effects in the 
$B^0$-$\overline{B^0}$ system.

Time-independent probabilities can further be obtained by integrating 
the rates ${\mit\Gamma}$:
$$
{\mit\Gamma}(f_1,f_2)={1\over\Gamma}\int_0^\infty d\tau\
{\mit\Gamma}(f_1,f_2;\tau)\ ;\eqno(37)
$$
this procedure gives the full rate for the double $B$-meson decay into
the final states $f_1$ and $f_2$.
Out of these quantities, one can form total dileptonic asymmetries:
$$
{\cal A}_{\ell\ell}={ {\mit\Gamma}(h^-,h^-)-{\mit\Gamma}(h^+,h^+)\over
{\mit\Gamma}(h^+,h^+)+{\mit\Gamma}(h^-,h^-) }\ ,\qquad
{\cal A}'_{\ell\ell}={ {\mit\Gamma}(h^+,h^-)-{\mit\Gamma}(h^-,h^+)\over
{\mit\Gamma}(h^+,h^+)+{\mit\Gamma}(h^-,h^-) }\ .\eqno(38)
$$
Unfortunately, these quantities, that are measured at $B$-factories,
are not sensitive to the non-standard parameters, because of the
suppression by factors $\delta$. For example, one has (to first
order in $\delta$):
$$
{\cal A}_{\ell\ell}=2\, {\cal R}e(\epsilon_S+\epsilon_L)
-4\, {\cal R}e(y_h)- {4\delta\over\omega^2}\, {\cal R}e(C)\ .\eqno(39)
$$
Instead, the asymmetries (38) can be used with confidence 
to constrain standard $CP$ and
$CPT$ violating effects, even in presence of dissipative phenomena.[33, 34]

More useful observables for identifying the presence of the extra,
non-standard parameters are the total $B^0$-$\overline{B^0}$ mixing
probability:
$$
\chi_B={ {\mit\Gamma}(h^+,h^+)+{\mit\Gamma}(h^-,h^-)\over
{\mit\Gamma}(h^+,h^+)+{\mit\Gamma}(h^-,h^-)+
{\mit\Gamma}(h^+,h^-)+{\mit\Gamma}(h^-,h^+)}\ ,\eqno(40)
$$
and the ratio of the total, same-sign to opposite-sign semileptonic 
rates:
$$
R_B={ {\mit\Gamma}(h^+,h^+)+{\mit\Gamma}(h^-,h^-)\over
{\mit\Gamma}(h^+,h^-)+{\mit\Gamma}(h^-,h^+)}\ .\eqno(41)
$$
They can both be expressed in terms of the following combination of
dissipative, non-standard parameters:
$$
X=A+\omega^2\, D+\omega\big[\omega\, {\cal I}m(B) - {\cal R}e(B)\big]\ ;
\eqno(42)
$$
in fact, one explicitly finds:
$$
\chi_B={\omega^2\over 2(1+\omega^2)}
\bigg[1+{2\over\omega^2(1+\omega^2)}\, X\bigg]
\ ,\qquad
R_B={\omega^2\over 2+\omega^2}\bigg[1+{4\over\omega^2(2+\omega^2)}\, X\bigg]
\ .\eqno(43)
$$
It should be stressed that while the dependence of $\chi_B$ and $R_B$
on the standard $CP$ and $CPT$ violating parameters is at most quadratic,[26]
the extra parameters responsible for possible dissipative extensions
of quantum mechanics appear linearly in (43), via the combination (42).
Therefore, these two observables are particularly suitable for checking
the presence of a non-vanishing value for the combination $X$.
Their measure provide two separate estimates for this quantity; instead,
in ordinary quantum mechanics $X=\,0$, thus $\chi_B$ and $R_B$ 
are not independent.

Unfortunately, so far both quantities in (43) have not been measured 
with great accuracy; $\chi_B$ is the better determined parameter
and the most recent data give: $\chi_B=0.156\pm0.024$.[32] 
Using this value and the already quoted one for
$\omega$, from (43) one gets the result: 
$X=(-4.5\pm6.3)\times 10^{-2}$.
In order to obtain estimates for the non-standard parameters in (5),
one needs measurements of other independent quantities besides $X$,
involving not only semileptonic final decay states. 
However, when $a=\,0$,[18] 
the form (5) of the extra piece in the equation (2) simplifies,
since complete positivity implies: 
$\alpha=\gamma$ and $b=c=\beta=\,0$.
In such a case, the previous result for $X$ gives a direct estimate
of the surviving parameter $\alpha$. Assimilating $1/\Gamma$ to the
$B^0$ lifetime, one finally gets: 
$\alpha=(-3.5\pm4.9)\times 10^{-14}\ {\rm GeV}$.
The accuracy on the determination of $X$ will be greatly improved by the
measurements at the $B$-factories. Indeed, preliminary estimates of
about one percent sensitivity to $R_B$ have been reported to be attainable
in [27], while a conservative estimate of two percent accuracy in the measure
of $\chi_B$ is indicated in [28]. If confirmed by the actual data, these
sensitivities will allow a determination of the combination $X$ with
about ten percent accuracy.

In conclusion, we have examined possible signals of the presence of
dissipative, non-standard dynamics in the evolution of the
$B^0$-$\overline{B^0}$ system. Various observables involving semileptonic
decays have been identified as being particularly sensitive to the
dissipative effects; these observables will be measured with great
accuracy in the new generation of dedicated $B$-experiments
(BaBar, Belle, CLEO-III, LHC-$b$). These new measurements,
besides being crucial for the confirmation of the standard model
theory of $CP$ violation, could very well provide clear evidence for
phenomena that go beyond ordinary quantum mechanics.

\vfill\eject

\centerline{\bf REFERENCES}
\bigskip

\item{1.} R. Alicki and K. Lendi, {\it Quantum Dynamical Semigroups and 
Applications}, Lect. Notes Phys. {\bf 286}, (Springer-Verlag, Berlin, 1987)
\smallskip
\item{2.} V. Gorini, A. Frigerio, M. Verri, A. Kossakowski and
E.C.G. Surdarshan, Rep. Math. Phys. {\bf 13} (1978) 149 
\smallskip
\item{3.} H. Spohn, Rev. Mod. Phys. {\bf 53} (1980) 569
\smallskip
\item{4.} W.H. Louisell, {\it Quantum Statistical Properties of Radiation},
(Wiley, New York, 1973)
\smallskip
\item{5.} C.W. Gardiner, {\it Quantum Noise} (Springer, Berlin, 1992)
\smallskip
\item{6.} M.O. Scully and M.S. Zubairy, 
{\it Quantum Optics} (Cambridge University Press, Cambridge, 1997)
\smallskip
\item{7.} L. Fonda, G.C. Ghirardi and A. Rimini, Rep. Prog. Phys.
{\bf 41} (1978) 587 
\smallskip
\item{8.} H. Nakazato, M. Namiki and S. Pascazio,
Int. J. Mod. Phys. {\bf B10} (1996) 247
\smallskip
\item{9.} F. Benatti and R. Floreanini, Phys. Lett. {\bf B428} (1998) 149
\smallskip
\item{10.} S. Hawking, Comm. Math. Phys. {\bf 87} (1983) 395; Phys. Rev. D
{\bf 37} (1988) 904; Phys. Rev. D {\bf 53} (1996) 3099;
S. Hawking and C. Hunter, Phys. Rev. D {\bf 59} (1999) 044025
\smallskip
\item{11.} J. Ellis, J.S. Hagelin, D.V. Nanopoulos and M. Srednicki,
Nucl. Phys. {\bf B241} (1984) 381; 
\smallskip
\item{12.} S. Coleman, Nucl. Phys. {\bf B307} (1988) 867
\smallskip
\item{13.} S.B. Giddings and A. Strominger, Nucl. Phys. {\bf B307} (1988) 854
\smallskip
\item{14.} M. Srednicki, Nucl. Phys. {\bf B410} (1993) 143
\smallskip
\item{15.} L.J. Garay, Phys. Rev. Lett. {\bf 80} (1998) 2508;
Phys. Rev. D {\bf 58} (1998) 124015
\smallskip
\item{16.} J. Ellis, N.E. Mavromatos and D.V. Nanopoulos, Phys. Lett.
{\bf B293} (1992) 37; Int. J. Mod. Phys. {\bf A11} (1996) 1489
\smallskip
\item{17.} F. Benatti and R. Floreanini, Ann. of Phys. {\bf 273} (1999) 58
\smallskip
\item{18.} F. Benatti and R. Floreanini, Phys. Lett. {\bf B451} (1999) 422
\smallskip
\item{19.} J. Ellis, J.L. Lopez, N.E. Mavromatos 
and D.V. Nanopoulos, Phys. Rev. D {\bf 53} (1996) 3846
\smallskip
\item{20.} P. Huet and M.E. Peskin, Nucl. Phys. {\bf B434} (1995) 3
\smallskip
\item{21.} F. Benatti and R. Floreanini, Nucl. Phys. {\bf B488} (1997) 335
\smallskip
\item{22.} F. Benatti and R. Floreanini, Phys. Lett. {\bf B401} (1997) 337
\smallskip
\item{23.} F. Benatti and R. Floreanini, Nucl. Phys. {\bf B511} (1998) 550
\smallskip
\item{24.} F. Benatti and R. Floreanini, 
Mod. Phys. Lett. {\bf A12} (1997) 1465; 
Banach Center Publications, {\bf 43} (1998) 71; 
Comment on ``Searching for evolutions 
of pure states into mixed states in the two-state system $K$-$\overline{K}$'',
{\tt hep-ph/9806450}
\smallskip
\item{25.} V.A. Kosteleck\'y and R. Van Kooten, Phys. Rev. D {\bf 54}
(1996) 5585
\smallskip
\item{26.} P. Colangelo and G. Corcella, Eur. Phys. J. C {\bf 1} 
(1998) 515
\smallskip
\item{27.} S. Yang and G. Isidori, Test of $CPT$ invariance in semileptonic
$B$ decays, BaBar \hbox{Note \#438}, 1998
\smallskip
\item{28.} A. Mohapatra, M. Satpathy, K. Abe and Y. Sakai,
Phys. Rev. D {\bf 58} (1998) 036003
\smallskip
\item{29.} The OPAL Collaboration, Zeit. fur Physik {\bf C76} (1997) 401
\smallskip
\item{30.} The OPAL Collaboration, CERN-EP/98-195, Eur. Phys. J. C (1999) 
\smallskip
\item{31.} N.W. Tanner and R.H. Dalitz, Ann. of Phys. {\bf 171} (1986) 463
\smallskip
\item{32.} Particle Data Group, Eur. Phys. J. C {\bf 3} (1998) 1
\smallskip
\item{33.} G. De Domenico and Ch. Y\`eche, Dilepton analysis in BaBar
experiment: measurement of the mixing parameter $\Delta m_B$ and study
of the T (CP) violation purely in mixing, BaBar Note \#409, 1998
\smallskip
\item{34.} The BaBar Collaboration, {\it The BaBar Physics Book}, 
SLAC-R-504, 1998, Ch. 11
\bye